\documentclass[]{amsart}
\begin{document}

\title[Conversion of dislocation oscillation waves to spin ones]
{Conversion of dislocation oscillation waves to spin ones in the
vicinity of OPT temperatures}

\author{D.I.Sirota(A)}
\address{(A) Bryansk State Technical University, 241035 Bryansk, Russia}

\author{A.F.Zhuravlyov (B)}
\address{(B) Taras Shevchenko Kyiv National University,
Physical Department, Glushkova 6, 03022 Kiev, Ukraine} \email{(B)
anjour@univ.kiev.ua}

\date{\today}

\begin{abstract}
Dislocation waves in magnetic crystals in the vicinity of
orientation phase transition (OPT) temperatures are considered in
a frame of the field theory of the defects. The singularities of
the dislocation flows, elastic deformations and magnetization
occur if the magnetic subsystem is inhomogeneous and the
dispersion of the media is not taken into account. Media
dispersion causes a regularity of these parameters and a
conversion of the spin wave to the dislocation wave.
\end{abstract}

\maketitle

\section{Introduction}

Real crystals contain defects of different nature, which
influence its wave properties. The dislocations influence of
phonon spectra of real crystals is especially notoriously.
Ferromagnetic structure is sensitive to the different influences
in the vicinity of the temperatures of Orientation Phase
Transitions (OPT). The low stability of the magnetic state of the
ferromagnetic near the temperature of OPT relieves an observation
of different non-linear effects. Therefore an interaction between
dislocations and magnetic structure of magnetics will be revealed
at these temperatures.

Phenomenological description of the dislocation ensembles is
based on the statistical effects. These effects are not a simple
sum of the properties of a number of dislocations. Statistical
properties of the dislocation ensembles reveal its wave
characteristics. One of these wave effects is a screening of
elastic deformations and correct description needs taking into
account dislocation cores.

Elastic fields of separated dislocations are screened therefore
the behavior of the dislocation core ensemble determinate the
interaction with the crystalline structure. Dislocation ensembles
move toward the direction of the influence in the case of
directed external influence. Crystal glides of a separate
dislocation of the ensemble realize a moving of the ensemble in
non-crystallographic directions \cite{1}.

\section{Basic Equations}

Quantitative description of the above model of the dislocation
ensemble is presented by the equations, which are analogous to
the Maxwell's equations \cite{1,2}:
\begin{equation}\label{1}
\nabla\times\hat{j}({\bf r},t)=\partial_t \hat{\alpha}({\bf r},t),
\end{equation}

\begin{equation}\label{2}
S\nabla\times\hat{\alpha}({\bf r},t)=-B\partial_t \hat{j}({\bf
r},t)-\hat{\sigma}({\bf r},t)
\end{equation}

Here $\hat{j}$ is the tensor of the density of the defect flow;
$\hat{\alpha}$ is the tensor of the defect density; $\hat{\sigma}$
is the tensor of the elastic deformations; $S$ is the constant
which describes the potential energy of the defect over the unity
of the length; $B$ is the constant, which describes the inertial
properties of the defect. In addition we take into account the
equation of the motion of the continuous media in the form:
\begin{equation}\label{3}
\rho\partial_t^2\hat{\sigma}({\bf r},t)=\int\hat{C}({\bf r},t;{\bf
r}_1 ,t_1 )\left (\nabla\bigotimes\nabla\hat{\sigma}({\bf
r},t)+\rho\frac{\partial\hat{j}({\bf r}_1 ,t_1 )}{\partial
t_1}\right )d{\bf r}_1 dt_1 ,
\end{equation}
where $\rho$ is the density; $\hat{C}({\bf r},t;{\bf r}_1 ,t_1 )$
is the elastic module tensor. We shall take into account the
interaction of the dislocation oscillation waves and the
magnetization oscillations putting the tensor $\hat{C}({\bf
r},t;{\bf r}_1 ,t_1 )$ as redefined one by magnetic-elastic
interaction \cite{3}. When the materials is characterized by
strong and intensive magnetic elastic interaction of the
dislocation oscillation waves and the magnetic moments, a
situation can results in the effects, which occur in the case when
the electric-magnetic waves spread in such materials
\cite{4}--\cite{5}. We point out that the redefined part of
$\hat{C}({\bf r},t;{\bf r}_1 ,t_1 )$ is proportional to dynamic
magnetic susceptibility tensor. Eq.\eqref{3} is
integer-differential equation due to the time and spatial
dispersions. Inhomogeneity of the considered material is
determinate by the inhomogeneity of the magnetic sub-system. It
may be easily achieved near the OPT temperatures by the
temperature gradient, or the external magnetic field. Below we
shall consider the cases when the sizes of the inhomogeneity along
the ${\bf 0x}$ axis are much more than the wavelengths of the
considered wave processes. In this case the inhomogeneity will be
take into consideration by the dependencies of the ferrimagnetic
resonance frequency on the coordinate $x$. We shall get in the
accordance with the Eq.\eqref{3}:
\begin{equation}\label{4}
\sigma=-\int\hat{G}({\bf r},t;{\bf r}_1 ,t_1 )\hat{C}({\bf r}_1
,t_1 ;{\bf r}_2 ,t_2 )\frac{\partial\hat{j}({\bf r}_2 ,t_2
)}{\partial t_2}d{\bf r}_1 dt_1 d{\bf r}_2 dt_2 ,
\end{equation}
Here $\hat{G}({\bf r},t;{\bf r}_1 ,t_1 )$ is the 4$^{th}$ rank
tensor of the Green's function of the Eq.\eqref{3}in the case
$\hat{j}({\bf r},t)=0.$

\section{Model simplification}

To simplify the model, let us consider screw component
oscillations plane wave, which is spread in the $z0x$ plane,
$\alpha_{yy}=\alpha$; ($\bf{0y}-$ axis is the high-symmetry axis).
In this case the Eqs.\eqref{1}, \eqref{2} take the form:
\begin{equation}\label{5}
\partial_t \alpha =\partial_z j_{xy}-\partial_x j_{zy},
\end{equation}

\begin{equation}\label{6}
B\partial_t j_{xy}=S\partial_z \alpha-\sigma_{xy},
\end{equation}

\begin{equation}\label{7} B\partial_t j_{zy}=-S\partial_z
\alpha-\sigma_{zy},
\end{equation}

In the absence of the magnetic-elastic interaction harmonic
dislocation wave is described by the dispersion equation:
\begin{equation}\label{8}
\omega^2 =\frac{1}{2B\rho}[\rho C_0 +(BC_0 +\rho S)k^2
\pm\sqrt{k^4 (\rho S-BC_0)^2 +2\rho C_0 (BC_0 +\rho S)k^2 +\rho^2
C_0^2}]
\end{equation}
Here the sign $(-)$ corresponds to the acoustic wave, and the sign
$(+)$ corresponds to the dislocation one. Thus the spectrum of the
oscillations of the dislocations has a gap $\omega_0$:
$$\omega_0 =\sqrt{\frac{C_0}{B}}.$$
Here $C_0=C^0_{xyxy}$ is the elasticity module. Measured gap value
and the value of Eq.\eqref{8} in the point $k=0$ enable us to
determine the constants $S$ and $B.$

\section{Dislocation--magnetization interaction}

Let us consider the interaction of the time-harmonic dislocation
wave with the magnetic subsystem. Neglecting of the spatial
dispersion and taking into account Eq.\eqref{4} and
$$G_{xyxy}=G_{zyzy}=\frac{1}{\rho\omega^2},$$ give us for long dislocation waves:
\begin{equation}\label{10}
\sigma_{xy}=-i\left
(\frac{C_0}{\omega}-\frac{4g\lambda^2\Omega}{M\omega (\Omega^2
-\omega^2 )}\right )j_{xy},
\end{equation}
\begin{equation}\label{11}
\sigma_{zy}=-i\left
(\frac{C_0}{\omega}-\frac{4g\lambda^2\Omega}{M\omega (\Omega^2
-\omega^2 )}\right )j_{zy}.
\end{equation}
Here $\omega$ is the dislocation wave frequency; $\lambda
=\lambda_{xyxy}=\lambda_{zyzy}$ is the magnetic-elastic constant;
$M$ is the sample magnetization, the easy axis is directed along
$\bf{0y}$ axis; $\Omega$ is the ferromagnetic resonance frequency,
which depends on the coordinate $x$; $g$ is the gyro-magnet ratio.
A substitution of the Eqs.\eqref{10}, \eqref{11} to the
Eqs.\eqref{6}, \eqref{7} gives us:
\begin{equation}\label{12}
S\partial_z\alpha =i\varepsilon (x)j_{xy},
\end{equation}
\begin{equation}\label{13}
-S\partial_x\alpha =i\varepsilon (x)j_{zy},
\end{equation}
where
\begin{equation}\label{14}
\varepsilon (x)=\omega B-\frac{1}{\omega}\left (C_0
-\frac{4g\lambda^2\Omega}{M(\Omega^2 -\omega^2 )}\right ).
\end{equation}
Then from the Eq.\eqref{5} we get:
\begin{equation}\label{15}
\frac{S}{\varepsilon (x)}\partial^2_z\alpha +S\partial_x \left (
\frac{1}{\varepsilon (x)}\partial_x\alpha \right )+\omega\alpha
=0.
\end{equation}
Due to the dependence of the frequency $\Omega$ (Eq.\eqref{14})
on the coordinate $x$ we can find a point $x_0 ,$ in which
$\varepsilon (x_0 )=0.$ In this point $$\Omega =\Omega_0
=-\frac{2g\lambda^2\omega_0}{(\omega^2 -\omega_0^2 )C_0
M}+\sqrt{\frac{4g^2 \lambda^4\omega_0^2}{(\omega^2 -\omega_0^2
)^2 C_0^2 M^2}+\omega^2}.$$ This point $x_0$ could be achieved in
materials with $\lambda\sim 10^8\frac{erg}{cm^3}, \Omega\sim 10^9
s^{-1}$ and $\omega\sim\omega_0\sim (\omega -\omega_0 )\sim\Omega
.$

Let us put $x_0 =0.$ In the vicinity of this point we can write:
\begin{equation}\label{17}
\varepsilon (x)=-\omega bBx,\qquad b>0.
\end{equation}
If a dependence of the ferromagnetic resonance frequency on the
coordinate $x$ caused by the temperature gradient in the
inhomogeneous media, we have $$b=-\frac{4g\lambda^2(\Omega_0^2
+\omega^2)}{MB\omega^2 (\Omega_0^2 -\omega^2
)^2}\frac{d\Omega_0}{dT}\frac{dT}{dx}.$$ Usually
$$\frac{d\Omega_0}{dT}<0,$$ therefore to get $b>0,$ we have put
$$\frac{dT}{dx}>0.$$

A substitution of the solution $$\alpha =\alpha_0 (x)e^{i(\omega
t-\kappa z)},$$ to the Eq.\eqref{15} and taking into account the
Eq.\eqref{17} gives us:
$$\partial_x(\frac{1}{x}\partial_x\alpha_0
(x))-\left ( \frac{\kappa^2}{x}+\frac{bB\omega^2}{S}\right
)\alpha_0 (x)=0.$$

Exact solution of this equation is complicated enough [6]. We are
interesting only in the part, which is connected with the
singularity [7]:
\begin{equation}\label{21}
\alpha_0 (x)=\alpha_0 (1+\frac{1}{2}\kappa^2 x^2{\rm ln}\kappa x).
\end{equation}
Therefore we have from this equation and the Eqs.\eqref{12},
\eqref{13}:
$$j_{xy}=\frac{\kappa\alpha_0 S}{bB\omega
x}e^{i(\omega t-\kappa z)},\qquad j_{zy}=i\frac{\kappa^2\alpha_0
S{\rm ln}\kappa x}{bB\omega}e^{i(\omega t-\kappa z)}.$$

Hence we got the singularities of the components of the
dislocation flow tensor and of the components of elastic stresses
tensor, as it follows from Eqs.\eqref{10}, \eqref{11}. The
singularities vanish if we shall take into account magnetization
oscillations attenuation: $$\Omega\Rightarrow\Omega
-i\delta\frac{\omega}{gM},$$ where $\delta$ is small attenuation
in magnetic subsystem \cite{8}. In the Eqs.\eqref{12}, \eqref{13}
we have substitute $\varepsilon (x)$ (Eq.\eqref{17})by:
$$\varepsilon (x)=-bB\omega x+iB\omega\nu .$$ Here $$\nu
=\frac{4\delta\lambda^2 (\omega^2 +\Omega_0^2 )}{\omega B\left [
(\Omega_0^2 -\omega^2 )^2 +\left ( \frac{2\omega\Omega_0\delta}{g
M}\right )^2\right ]^2 M^2}.$$

Let us calculate averaged magnetic-elastic energy which is
absorbed in the vicinity of the point $x=0:$
$$Q=\frac{i\omega}{2}\int\overline{\Re\sigma_{xy}\Re\varepsilon_{xy}}dx.$$
Here averaging is accomplished on the oscillation period;
$\varepsilon_{xy}$ is the component of the elastic deformation
tensor.

So we get from the Eq.\eqref{10}:$$Q=-\frac{1}{2\omega}\left
(\frac{S\kappa\alpha_0}{B}\right )^2\int\frac{\nu dx}{(bx)^2
+\nu^2}.$$ If the value of $\nu$ is small, we can write:
$$\frac{\nu}{(bx)^2 +\nu^2}\Rightarrow\delta (bx).$$

Hence energy absorption is resonance-like, realized in the narrow
area in the vicinity of the point $x=0,$ where the dislocation
wave frequency coincides with the gap which is redefined by the
magnetic-elastic interaction in the dislocation oscillations
spectrum. Finally we got: $$Q=-\frac{1}{2b\omega}\left
(\frac{S\kappa\alpha_0}{B}\right )^2 .$$

The velocity of the absorption of the dislocation oscillation
energy depends not on the value of $\nu$ because the effective
width of the region of intensive energy absorption is
proportional to $\nu .$

A conversion of the dislocation waves to the spin ones may be
realized due to the interaction of these waves with the
magnetization oscillations. The space dispersion influence on
this process is essential. Let us consider this effect. The value
of the magnetic field, which arise due to the elastic deformation
is given by $$H_x
=-\frac{\lambda\varepsilon_{xy}}{M}=2i\frac{\lambda\omega\rho}{M}G(\omega
)j_{xy},$$ in the absence of the dispersion. The magnetization we
can write as: $$M_x =\chi_{xx}H_x
=ig\frac{S\lambda\kappa\alpha_0\Omega_0}{bx\omega^2 (\Omega_0^2
-\omega^2 )}e^{i(\omega t-\kappa z)},$$

$$M_z =\chi_{zx}H_x
=-g\frac{S\lambda\kappa\alpha_0}{bx\omega (\Omega_0^2 -\omega^2
)}e^{i(\omega t-\kappa z)}.$$ Here $\chi_{xx}$ and $\chi_{zx}$ are
the components of the tensor of magnetic susceptibility.

One can see that in the point $x=0$ magnetisation has a
singularity. To take into account spin wave dispersion we have
substitute in the Eqs.\eqref{12}, \eqref{13} $$k^2\Rightarrow
-\partial^2_x ,$$ where $k$ is the $x$ component of the wave
vector. Then we get for the spin waves
\begin{equation}\label{34}
\beta\partial_x^2 M_x -bxM_x =igS\alpha_0
\frac{\lambda\kappa\Omega_0}{B(\Omega_0^2 -\omega^2
)\omega^2}e^{i(\omega t-\kappa z)},
\end{equation}
\begin{equation}\label{35}
\beta\partial_x^2 M_z -bxM_z =-gS\alpha_0
\frac{\lambda\kappa}{B(\Omega_0^2 -\omega^2 )\omega}e^{i(\omega
t-\kappa z)}.
\end{equation}
The values of $\chi_{xx}$ and $\chi_{zx}$ are taken in the point
$x=0$ in these equations, $$\beta =2gMC_0 \gamma\frac{\Omega_0
(\omega^2 -\omega_0^2 )}{B(\omega^2
-\Omega_0^2)\omega^2\omega_0^2}>0,$$ where $\gamma$ is the
inhomogeneous exchange parameter.

\section{Results and discussion}
The solution of the Eqs.\eqref{34}, \eqref{35} describes spin wave
which is created by dislocation one. This spin wave moves from the
point $x=0:$
$$M_x =-\frac{gS\alpha_0\lambda\kappa\Omega_0}{B(\Omega_0^2 -\omega^2
)\omega^2 b^{2/3}\beta^{1/3}}e^{i(\omega t-\kappa z)}\int_0^\infty
d\nu{\rm exp}\left \{i\left [x\nu\left (\frac{b}{\beta}\right
)^{1/3}+\frac{(\nu )^3}{3}\right ]\right \},
$$
$$M_z =-i\frac{gS\alpha_0\lambda\kappa}{B(\Omega_0^2 -\omega^2
)\omega b^{2/3}\beta^{1/3}}e^{i(\omega t-\kappa z)}\int_0^\infty
d\nu{\rm exp}\left \{i\left [x\nu\left (\frac{b}{\beta}\right
)^{1/3}+\frac{(\nu )^3}{3}\right ]\right \}.
$$

In the region $|x|>>\lambda,\quad x<0$ we can estimate the values
of $M_x$ and $M_z$ as:
$$M_x =\frac{-i\Omega_0}{\omega}M_z =\frac{gS\alpha_0\lambda\kappa\Omega_0\sqrt{\pi}}{B(\Omega_0^2 -\omega^2
)\omega b^{3/4}\beta^4 x^{1/4}}{\rm exp}\left \{i\left (\omega
t-\kappa
z-\frac{2}{3}\sqrt{\frac{b|x|^3}{\beta}}-\frac{\pi}{4}\right
)\right \}.$$

This effect takes place in the case when the dislocation wave is
directed at some angle $\theta$ to the $\bf{0x}$ axis. We shall
estimate the value of $\theta$ when the considered conversion is
maximal. In the approximation of the geometrical optic a
wavelength of dislocation wave is much less than the size of the
inhomogeneity
\begin{equation}\label{40}
\sqrt{(\omega^2 -\omega_0^2)\frac{B}{S}}>>b
\end{equation}
to the left-hand direction of the reflection point $x_0$ which is
determined by:
\begin{equation}\label{41}
S\kappa^2 +Bb\omega^2 x_0 =0,
\end{equation}
and the amplitude $\alpha$ of the oscillations decreases
exponentially. The point $x=0$ is situated in the right-hand
direction from the point $x_0.$ The observation of this effect is
possible in the case when the points $x=0$ and $x_0$ are in
proximity one to another:
\begin{equation}\label{42}
|kx_0|\approx 1,
\end{equation}
where $k$ is the $x-$component of the wave vector.

At the large distance from the reflection point a magnetic-static
influence is negligible, and the dispersion equation is:
$$k^2 +\kappa^2 =\frac{B}{S}(\omega^2 -\omega_0^2 ).$$
Hence we have $$\kappa^2 =\frac{B}{S}(\omega^2 -\omega_0^2 ){\rm
sin}^2\theta .$$ It is easy to get in the vicinity of the point
$x=0$ $$k^2 +\kappa^2 =-\frac{Bb\omega^2}{S}x,$$ and
$$|k|\approx\kappa =\sqrt{\frac{B}{S}(\omega^2 -\omega_0^2 ){\rm sin}\theta}.$$
We substitute this equation to the equation for $x_0,$ which
follows from Eq.\eqref{41}, then to the Eq.\eqref{42} and take
into account Eq.\eqref{40} to get $${\rm
sin}^3\theta\approx\sqrt{\frac{S}{B}}\frac{\omega^2}{(\omega^2
-\omega_0^2 )^{3/2}}b<<1.$$ Here we can see that the angle
$\theta$ is small but is not equal to zero.

\end{document}